\documentclass{PoS}

\usepackage{amsfonts}
\usepackage{amsmath}
\usepackage{subfigure} 

\graphicspath{{./figs/}}

\title{
  Continuum extrapolation of $B_K$ with staggered fermions
}

\ShortTitle{
  Continuum extrapolation of $B_K$
}

\author{\speaker{Weonjong Lee}, 
  Yong-Chull Jang, Hyung-Jin Kim, Jangho Kim,
  Kwangwoo Kim, Boram Yoon \\
  Lattice Gauge Theory Research Center, CTP, and FPRD, \\
  Department of Physics and Astronomy,
  Seoul National University, Seoul, 151-747, South Korea \\
  E-mail: \email{wlee@snu.ac.kr}}

\author{Taegil Bae \\
  Korea Institute of Science and Technology Information, 
  Daejeon, 305-806, South Korea \\
  E-mail: \email{esrevinu@gmail.com}}

\author{Chulwoo Jung \\
  Physics Department, Brookhaven National Laboratory,
  Upton, NY11973, USA \\
  E-mail: \email{chulwoo@bnl.gov}}

\author{Jongjeong Kim \\  
  Physics Department,
  University of Arizona,
  Tucson, AZ 85721, USA \\
  E-mail: \email{rvanguard@gmail.com}}

\author{Stephen R. Sharpe\\
  Physics Department, University of Washington, 
  Seattle, WA 98195-1560, USA \\
  E-mail: \email{sharpe@phys.washington.edu}}

\author{SWME Collaboration}

\abstract{We report on recent progress in the calculation of $B_K$ using
  HYP-smeared staggered fermions on the MILC asqtad lattices.
  Our main focus is on the continuum extrapolation, which is done using
  (up to) four different lattice spacings---$a\approx$ 0.12, 
  0.09, 0.06 and 0.045 fm.
  Since Lattice 2010, we have reduced the statistical errors
  on the $a\approx 0.09\;$fm lattices by a factor of $\sim 3$, and
roughly doubled the size of the $a\approx0.045\;$fm ensemble.
  We find that these improvements have a very significant impact on
the continuum extrapolation, with the $a\approx 0.12\;$fm data lying
outside the range of applicability of simple functional forms.
Hence we use only the three smallest lattice spacings to perform the
extrapolation, finding
$\hat{B}_K = B_K(\text{RGI}) = 0.725 \pm 0.004(\text{stat}) \pm
  0.038(\text{sys}) $. This value is consistent with our published
value from 2010 (based the three coarsest lattice spacings), 
but has smaller errors.}

\FullConference{
The XXIX International Symposium on Lattice Field Theory - Lattice 2011\\
July 10-16, 2011\\
Squaw Valley, Lake Tahoe, California
}

\begin{document}

\section{Introduction} 

The calculation of the kaon mixing matrix element $B_K$ is
one of the successes of lattice QCD. Results with all errors
controlled are available with different fermion discretizations,
the most recent entry being that using Wilson fermions~\cite{BMWBK}. 
For a summary, see Refs.~\cite{FLAG} and~\cite{LatticeAverages}.
There is some tension between these results, which needs to be resolved
in order to know whether the Standard Model can describe CP-violation
in kaon mixing. 

In this proceedings we update our results for $B_K$.
These are obtained using improved staggered fermions,
specifically HYP-smeared valence quarks on asqtad sea-quarks. 
We describe here results obtained with chiral fitting functions
from SU(2) staggered chiral perturbation
theory (SChPT), which give our most reliable results~\cite{BKPRD}.
We compare our results to those obtained a year ago and presented
at Lattice 2010~\cite{ref:wlee-2010-1}.
In companion proceedings we update the analysis based on
SU(3) SChPT~\cite{Lat2011SU3} and discuss strategies
for dealing with correlations in chiral fits~\cite{Lat2011Fitting}.

Table \ref{tab:milc-lat} lists all the ensembles on which
we have calculated $B_K$, and notes which results have changed
in the last year. In particular, since Lattice 2010 we have
accumulated higher statistics on 
the C2, C5, F1, U1 ensembles and added a new measurement on S2.
Of these, the most important updates are those on
F1 and U1, since they are used (together with C3 and S1) to
do our continuum extrapolation. We focus on this issue here.
The other updates, as well as the yet unanalyzed ensemble S2,
provide information on the sea-quark mass dependence at multiple
lattice spacings. 
%
%

%
%
%
\begin{table}[h!]
\begin{center}
\begin{tabular}{ c | c | c | c | c | c | c }
\hline
$a$ (fm) & $am_l/am_s$ & geometry & ID & ens $\times$ meas 
& $B_K$($\mu=2$ GeV) & status \\
\hline
0.12 & 0.03/0.05  & $20^3 \times 64$ & C1 & $564 \times 1$ &  0.556(14) & old \\
0.12 & 0.02/0.05  & $20^3 \times 64$ & C2 & $486 \times 9$ &  0.568(16) & \texttt{update} \\
0.12 & 0.01/0.05  & $20^3 \times 64$ & C3 & $671 \times 9$ &  0.566(5)  & old \\
0.12 & 0.01/0.05  & $28^3 \times 64$ & C3-2 & $275 \times 8$ &  0.570(4)  & old \\
0.12 & 0.007/0.05 & $20^3 \times 64$ & C4 & $651 \times 10$ & 0.563(5)  & old \\
0.12 & 0.005/0.05 & $24^3 \times 64$ & C5 & $509 \times 9$ &  0.566(4) & \texttt{update} \\
\hline
0.09 & 0.0062/0.031 & $28^3 \times 96$ & F1 & $995 \times 9$ & 0.527(4) & \texttt{update} \\
0.09 & 0.0031/0.031 & $40^3 \times 96$ & F2 & $850 \times 1$ & 0.551(9) & \texttt{update} \\
\hline
0.06 & 0.0036/0.018 & $48^3 \times 144$ & S1 & $744 \times 2$ & 0.537(7) & old \\
0.06 & 0.0025/0.018 & $56^3 \times 144$ & S2 & $198 \times 9$ & -NA- & \texttt{new} \\
\hline
0.045 & 0.0028/0.014 & $64^3 \times 192$ & U1 & $705 \times 1$ & 0.530(7) & \texttt{update} \\
\hline
\end{tabular}
\end{center}
\caption{MILC asqtad ensembles used to calculate $B_K$. 
$a m_\ell$ and $a m_s$ are the masses, in lattice units,
of the light and strange sea quarks, respectively. ``ens'' indicates
the number of configurations on which ``meas'' measurements are made.
The results for $B_K(\text{NDR}, 2 \text{ GeV})$ are obtained
using the SU(2) SChPT fits (4X3Y-NNLO) discussed in the
text.
}
\label{tab:milc-lat}
\end{table}

\section{Chiral fits}
In our numerical study, our lattice kaons are composed of valence
(anti)quarks with masses $m_x$ and $m_y$. 
On each MILC ensemble, we use 10 valence masses:
\begin{equation}
  am_x, am_y = am_s \times {n}/{10} \qquad \text{with} \qquad 
  n = 1,2,3,\ldots,10,
\end{equation}
where $am_s$ is the strange sea quark mass.
In our standard fits we extrapolate to $am_d^{\rm phys}$
using the lowest 4 values for $am_x$ (the ``X-fit''---done
at fixed $am_y$),
and then extrapolate to $m^\textrm{phys}_s$ using the
highest 3 values of $am_y$ (``Y-fit'').
As described in Ref.~\cite{BKPRD}, these choices keep us
in the regime where we expect next-to-leading order (NLO)
SU(2) ChPT to be reasonably accurate.
The X-fits described here are done to the form predicted by
NLO partially quenched SChPT (and given in Ref.~\cite{BKPRD}), 
augmented by a single analytic term of next-to-next-to-leading order (NNLO).
The $am_y$ dependence (which is not controlled by ChPT)
is very close to linear and we use a linear fit for our central values.
We dub this entire fitting procedure the ``4X3Y-NNLO fit'', and use
it for our central values. Other choices (e.g. NLO vs. NNLO) are
used to estimate fitting systematics.

\begin{figure}[t!]
\centering
\subfigure[Lattice 2010]{\includegraphics[width=0.49\textwidth]
  {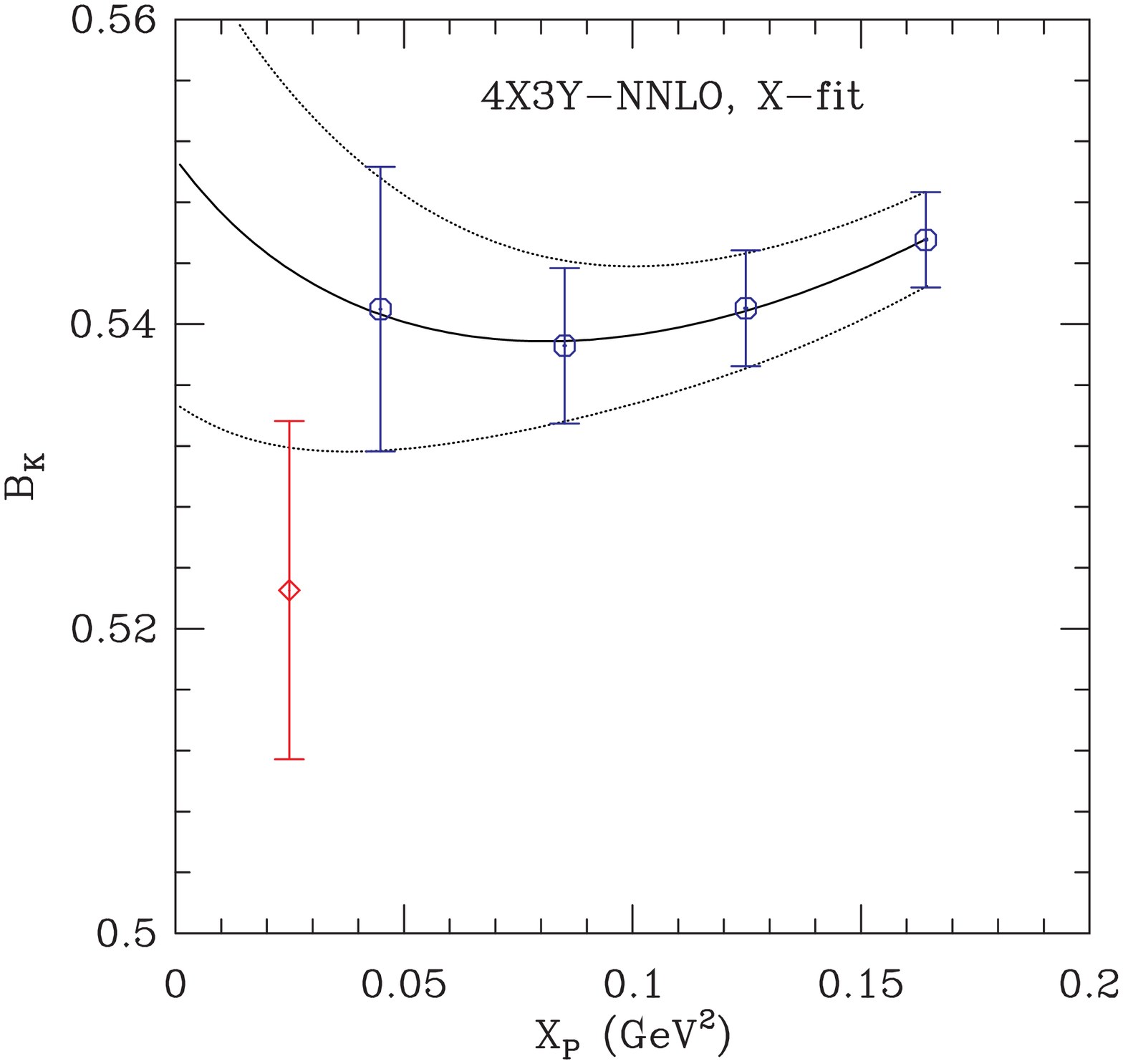}}
\subfigure[Lattice 2011]{\includegraphics[width=0.49\textwidth]
  {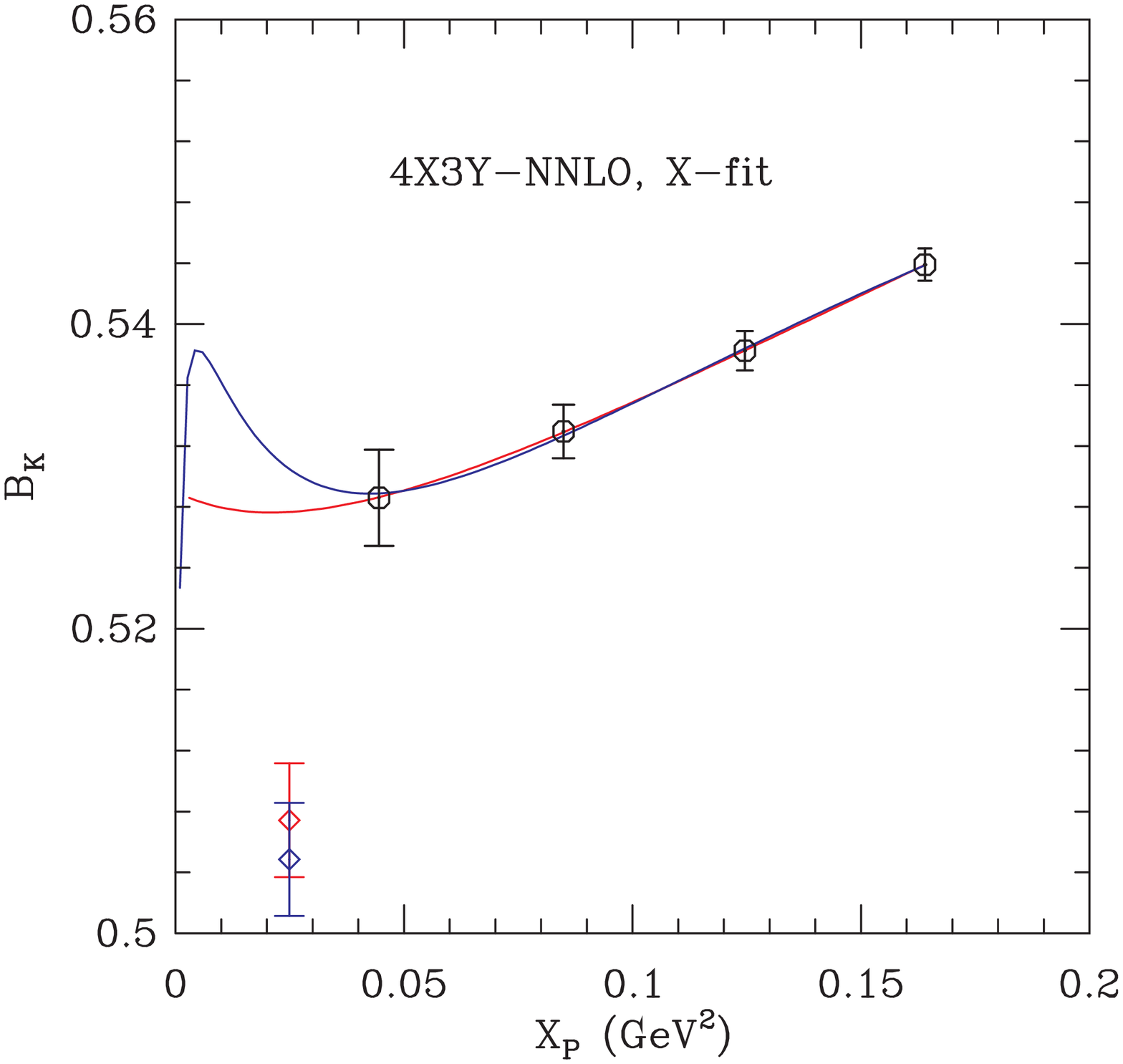}}
\caption{X-fits to $B_K(1/a)$ for the F1 ensemble,
  with 1 measurement/configuration (left) and with 9 measurements/configuration 
  (right). $X_P$ is the mass-squared of the $\bar x x$ valence pion,
  and we fix $am_y = 0.031$ (the heaviest value). In the right panel, the
  blue curve includes finite-volume (FV) corrections, 
  while the red curve does not. FV corrections are not included in the
  left panel. Extrapolated results (diamonds) have
  known taste-breaking discretization errors removed 
and thus do not lie on the curves.}
\label{fig:su2-4x3y-nnlo:F1}
\end{figure}
%
%
%
Examples of the X-fits are shown in Fig.~\ref{fig:su2-4x3y-nnlo:F1}.
Here we compare results on the F1 ensemble from Lattice 2010 with those
from our present dataset. 
In each panel, a fit to the SChPT form is shown,
%
%
together with the result after extrapolating
$a m_x\to a m_d^{\rm phys}$ and removing known discretization errors.
For details of this procedure, see Ref.~\cite{BKPRD}.

The addition of 8 extra measurements per configuration
reduces the statistical errors by $\sim 3$, as expected.
In addition, the central values shift down by about $1\sigma$,
leading to a correspondingly lower value of $B_K$.
We have also now incorporated finite volume corrections,
as predicted by NLO SChPT,
into the fitting function~\cite{ref:wlee-2011-1}.
\section{Continuum Extrapolation}
\begin{figure}[t!]
\centering
\subfigure[Lattice 2010]{\includegraphics[width=0.49\textwidth]
  {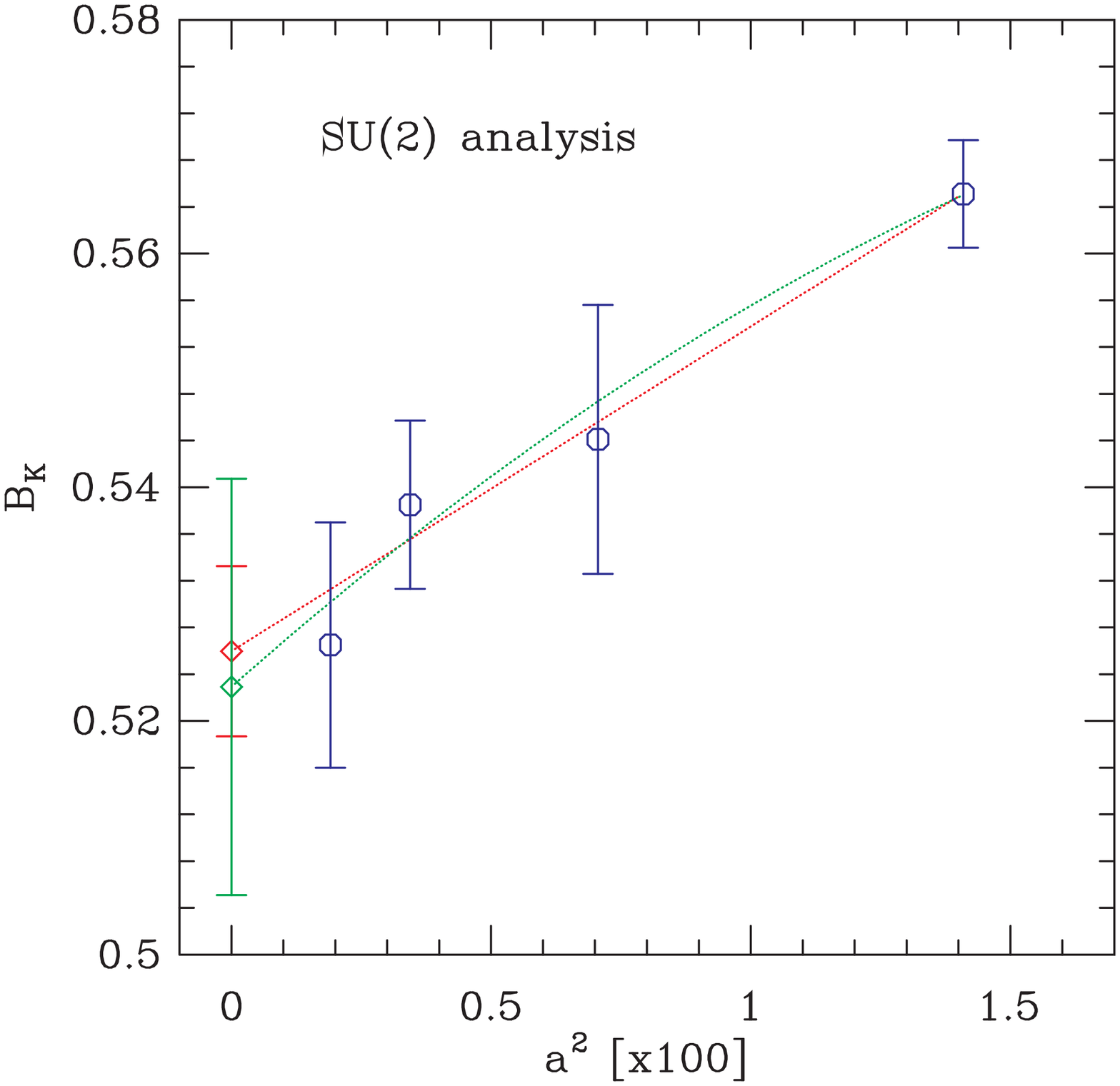}}
\subfigure[Lattice 2011]{\includegraphics[width=0.49\textwidth]
  {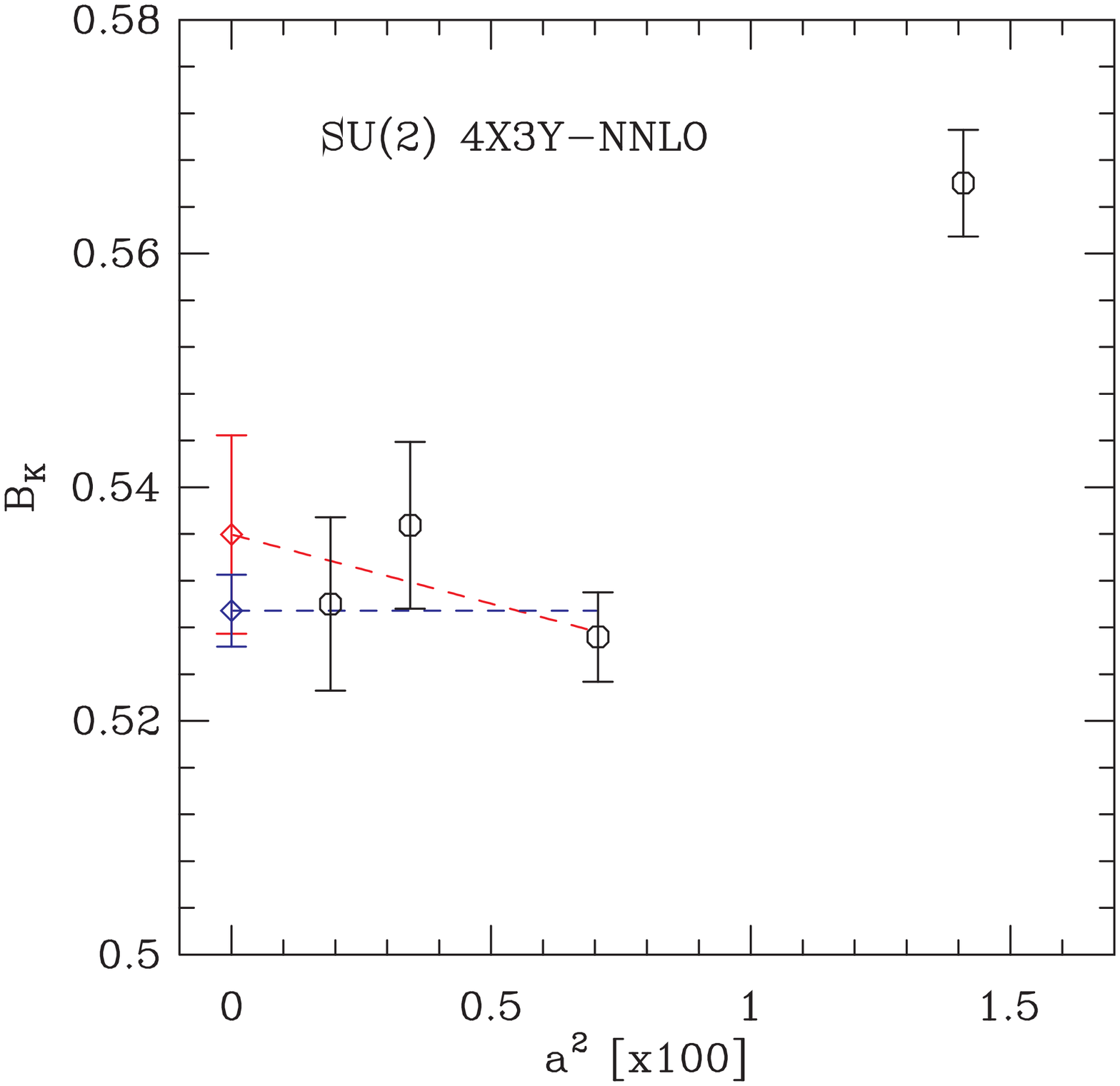}}
\caption{Continuum extrapolation of $B_K(\mu=2\;{\rm GeV})$ 
 versus $a^2$ (in units of $0.01\;{\rm fm}^2$). The left panel
shows results from Lattice 2010, with linear and quadratic fits.
The right panel shows our updated results, with constant and linear
fits to the leftmost three points.}
\label{fig:su2-4x3y-nnlo:scale}
\end{figure}
We do our continuum extrapolation using the four lattices with
$a m_\ell/a m_s=0.2$. Figure~\ref{fig:su2-4x3y-nnlo:scale} shows
how this extrapolation is changed by our updated results.
Clearly, the most significant change is for the F1 ensemble,
where the downward shift rules out the possibility of a straightforward
extrapolation using all four points. We thus first
consider extrapolations based only on the
smallest three values of $a$. The figure shows results both for a constant
and a linear fit, which have good
$\chi^2/d.o.f$ as listed in Table~\ref{tab:fit-func}.

\begin{figure}[tb!]
\centering
\subfigure[Unconstrained]{\includegraphics[width=0.49\textwidth]
  {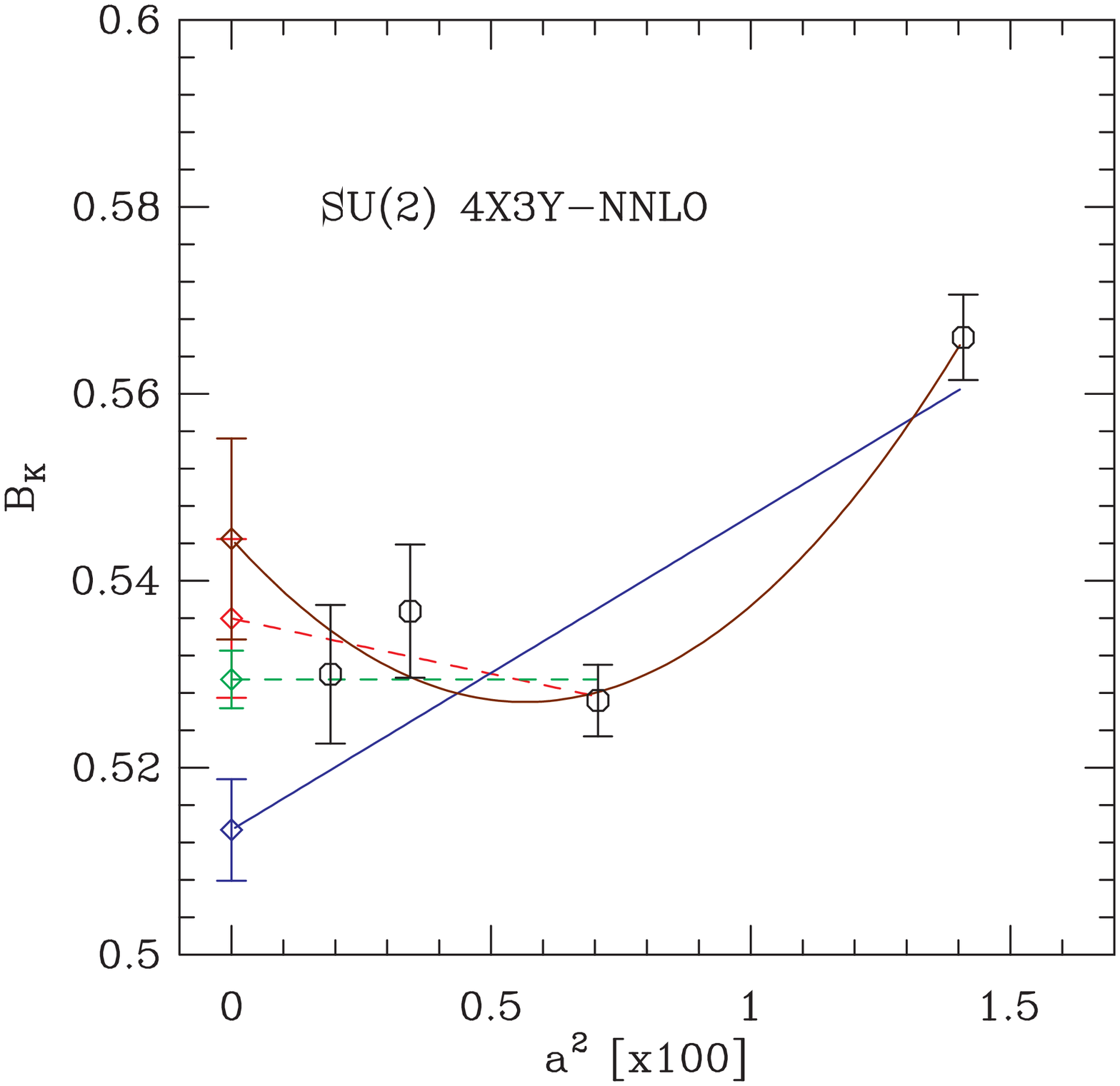}}
\subfigure[Bayesian]{\includegraphics[width=0.49\textwidth]
  {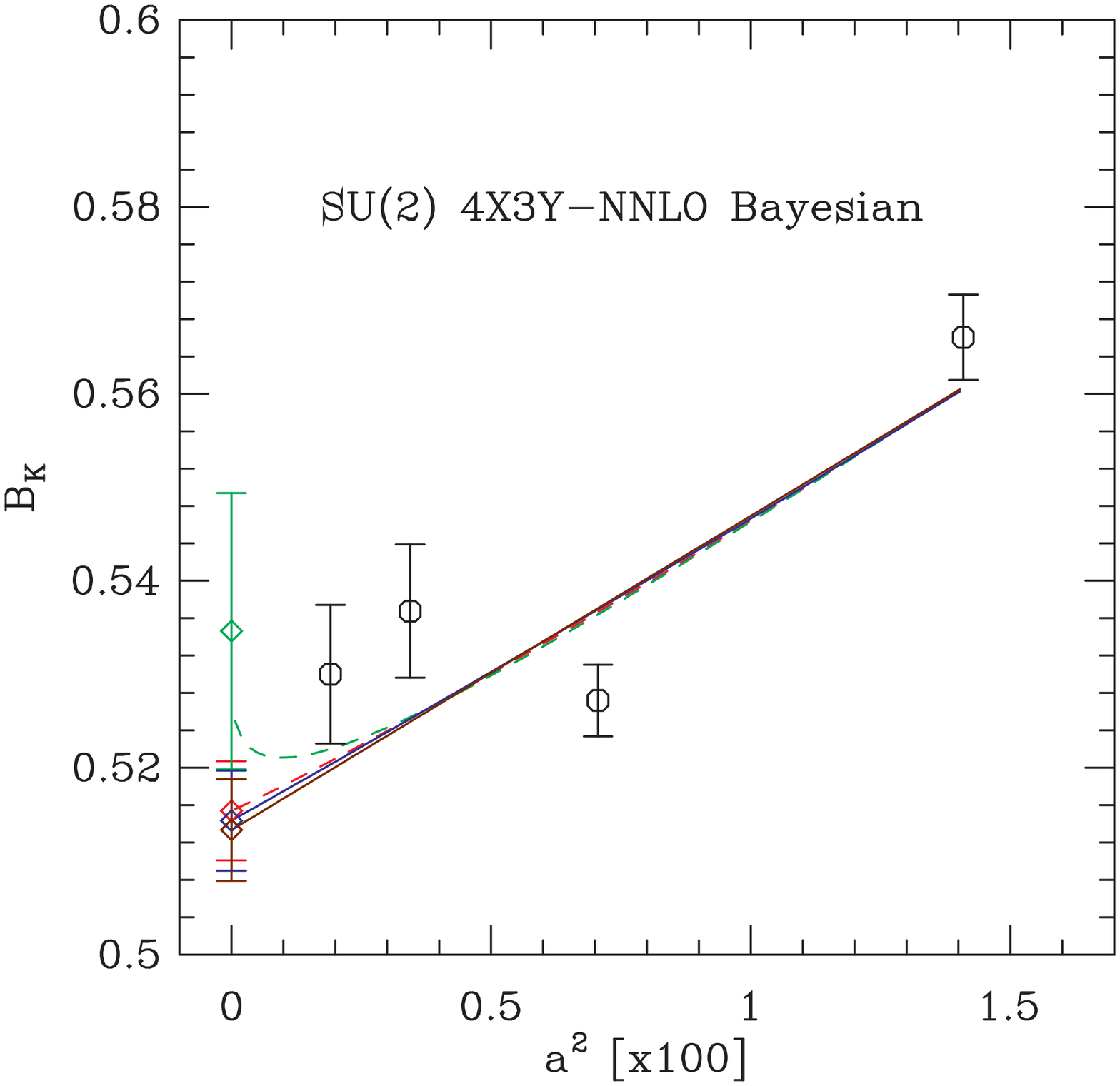}}
\caption{Continuum extrapolations of $B_K(\mu=2\;{\rm GeV})$ 
with different fitting choices. Fits are described in the text.}
\label{fig:su2:scale}
\end{figure}
\begin{table}[bth!]
\begin{center}
\begin{tabular}{c | c | l | c | c}
\hline
\hline
fit type & \# data & fit function & $\chi^2$/d.o.f & $\chi^2$/d.o.f. 
(Bayes)\\
\hline
{\tt const}  & 3      & $ c_1 $ & 0.70 & -NA-\\
{\tt lin3}   & 3      & $ c_1 + c_2 a^2 $ & 0.73 & -NA- \\
{\tt lin4}   & 4      & $ c_1 + c_2 a^2 $ & 6.35 & -NA- \\
{\tt quad}   & 4      & $ c_1 + c_2 a^2 + c_3 a^4 $ & 1.44 & 4.39 \\
{\tt a2a2g2} & 4      & $ c_1 + c_2 a^2 + c_3 a^2 \alpha_s(a) $ 
 & 1.66 & 4.16 \\
{\tt a2g4}   & 4      & $ c_1 + c_2 a^2 + c_3 \alpha_s^2(a) $ & 3.49 & 4.48 \\
{\tt a2a2g2g4a4} & 4  & $ c_1 + c_2 a^2 + c_3 a^2 \alpha_s(a) 
 + c_4 \alpha_s^2(a) + c_5 a^4 $ & -NA- & 3.86 \\
\hline
\hline
\end{tabular}
\end{center}
\caption{Functional forms used for continuum extrapolations. 
Fits to 3 data points exclude the largest value of $a$.
Bayesian fits and $\chi^2$ are described in the text.}
\label{tab:fit-func}
\end{table}

We now describe our attempts to fit
all four points to a reasonable functional form.
As explained in Ref.~\cite{ref:VdWS}, the expected dependence
is a linear combination of $a^2$, $a^2\alpha_s(1/a)$
and $\alpha_s^2(1/a)$, together with higher order terms.
Given that we have only four points, we began by using
three-parameter fitting functions, in particular those
labeled 
{\tt quad}, {\tt a2g4} and {\tt a2a2g2} 
in Table~\ref{tab:fit-func}, in addition to the two-parameter linear
fit {\tt lin4}.
Examples of the resulting fits are shown in Fig.~\ref{fig:su2:scale}.
The left panel shows unconstrained fits to all points using
the linear ({\tt lin4}--blue curve) 
and quadratic ({\tt quad}--brown curve) functions,
as well as reproducing the constant and linear fits to the three
smallest values of $a$ from Fig.~\ref{fig:su2-4x3y-nnlo:scale}.
Both the {\tt lin4} and {\tt quad} fits are problematic.
The former has a poor $\chi^2$ (see Table~\ref{tab:fit-func}),
although the value of the coefficient 
$c_2 \approx (0.36\; \text{GeV})^2$
is reasonable.
The quadratic fit has a reasonable $\chi^2$,
but has unphysically large values for the coefficients.
In particular,
we expect $|c_2| = \Lambda_2^2$ and $|c_3|= \Lambda_3^4$ with 
$\Lambda_2\approx\Lambda_3\approx\Lambda_{\rm QCD} \approx 0.3\;\text{GeV}$,
while the fit gives $\Lambda_2 \approx 0.49\; \text{GeV}$
and $\Lambda_3\approx 0.95\;$GeV. The latter value is
unreasonably large.
Similar problems occur for the {\tt a2a2g2} and {\tt a2g4} fit forms.
%

Thus we have repeated the {\tt a2a2g2}, {\tt a2g4}
and {\tt quad} fits imposing Bayesian constraints.
We augment the $\chi^2$ in the usual way, with the expected
central values of the coefficients $c_2$ and $c_3$ being zero,
while the expected standard deviations are 
\begin{eqnarray}
\sigma_{c_2} &=& 2 \Lambda^2 \ \ 
\textrm{for {\tt a2a2g2}, {\tt a2g4} \& {\tt quad},} 
\\ 
\sigma_{c_3} &=& (2 \Lambda^2),\ 2\ \&\ (2 \Lambda^4) \ \ 
\textrm{for {\tt a2a2g2}, {\tt a2g4} \& {\tt quad},} 
\end{eqnarray}
with $\Lambda= 300\;$MeV.
The resulting fits are shown in 
Fig.~\ref{fig:su2:scale} (b), with the augmented
$\chi^2$/d.o.f given in Table~\ref{tab:fit-func}.
The red curve shows the \texttt{a2a2g2} fit,
the green curve the \texttt{a2g4} fit and
the blue curve the \texttt{quad} fit.
For reference we also show the unconstrained \texttt{lin4} fit
in brown. 
We find that all three constrained fits are poor,
with the {\tt a2a2g2} and {\tt quad} fits differing little
from the {\tt lin4} fit.

We have also attempted a fit to 
$c_1+c_2 a^2+c_3 a^2 \alpha_s(a)
+c_4 \alpha_s(a)^2 + c_5 a^4$, 
which is possible as long as one uses Bayesian constraints
on $c_{2-5}$. 
The resulting fit, shown by the black curve in
Fig.~\ref{fig:su2:scale} (b), remains poor,
with $\chi^2/d.o.f.$ given in Table~\ref{tab:fit-func}.
Thus we conclude that we cannot describe the data from
all four values of $a$ with any of our fit functions
if we insist on physically reasonable coefficients.
Most likely this indicates that more terms are needed in
the fit functions.

Fortunately, the uncertainty in the correct global fit
form is not that important for the value in the continuum limit.
We take the result from the {\tt const} fit 
for our central value, 
and use the difference
with the Bayesian {\tt a2a2g2} fit for an extrapolation systematic.
As Fig.~\ref{fig:su2:scale} (b) shows, we would obtain
essentially the same systematic error were we to
use either of the other two Bayesian fits or the {\tt lin4} fit.

\section{Updated Result and Error Budget}

Our updated result for $B_K$ from the SU(2)-SChPT analysis is
\begin{equation}
\hat{B}_K = B_K(\text{RGI}) = 
0.725 \pm 0.004(\text{stat}) \pm 0.038(\text{sys}) \,.
\end{equation}
Here we use the 4X3Y-NNLO fit to valence quark mass dependence 
and the {\tt const} fit for the continuum extrapolation.

%
%
\begin{table}[htb!]
\centering
\begin{tabular}{ l | l l r }
\hline \hline
cause & error (\%) & memo & status \\
\hline
statistics      & 0.58   & 4X3Y-NNLO fit $+$ {\tt const} & update\\
matching factor & 4.4    & $\Delta B_K^{(2)}$ (U1) & \cite{ref:wlee-2010-1} \\
discretization  & 2.7    & diff.~of const and a2a2g2 (Bayesian) & update \\
fitting (1)     & 0.92   & X-fit (C3) & \cite{BKPRD}\\
fitting (2)     & 0.08   & Y-fit (C3) & \cite{BKPRD}\\
$am_l$ extrap   & 0.06   & diff.~of (C3) and linear extrap 
                                      & \cite{BKPRD}\\
$am_s$ extrap   & 0.5    & diff.~of constant vs linear extrap 
                                      & \cite{BKPRD}\\
finite volume   & 0.59   & diff.~of $V=\infty$ and FV fits 
                          & update~\cite{ref:wlee-2011-1} \\
$r_1$           & 0.14   & $r_1$ error propagation (C3) 
                                      & \cite{BKPRD}\\
$f_\pi$         & 0.38   & 132 MeV vs. 124.4 MeV 
                                      & \cite{BKPRD}\\
\hline \hline
\end{tabular}
\caption{Error budget for $B_K(2\;{\rm GeV})$ obtained using 
  SU(2) SChPT fitting.
The ``memo'' column gives a brief description of the error estimate,
while ``status'' indicates whether the error is updated 
or the same as in Lattice 2010~\cite{ref:wlee-2010-1} or our
paper~\cite{BKPRD}.
  \label{tab:su2:err-budget}}
\end{table}
The error budget for this result is given in Table \ref{tab:su2:err-budget}.
Most of the errors are estimated as in Ref.~\cite{BKPRD}, 
and many are unchanged from that work, as indicated in the ``status'' column.
The major changes in the last year are in the statistical
and discretization errors.
The former has been substantially reduced
compared to the 1.4\% quoted at Lattice 2010~\cite{ref:wlee-2010-1}.
By contrast, the discretization error has substantially
increased (from 0.1\%), because of our poorer understanding
of the continuum extrapolation. Previously we used the
difference between the result on the U1 ensemble and the
continuum value as an estimate of this error, while here
we use the difference between the {\tt const} and {\tt a2a2g2} fits.

We have also changed our method of calculating finite volume errors,
but this has a minor impact on the final error.

Combining the statistical and systematic errors in quadrature,
our updated result has a 5.3\% error. This is slightly larger
than the 4.8\% error in our result from Lattice 2010
($\hat{B}_K=0.720\pm 0.010 \pm 0.033$~\cite{ref:wlee-2010-1}).
The increase is due to the larger discretization systematic,
which overwhelms the reduction in the statistical error.
Still, the overall error is changed very little, because our
dominant systematic remains the truncation error introduced
by our use of 1-loop perturbative operator matching.\footnote{%
Our updated error is smaller than that in our published result
($\hat{B}_K=0.724\pm 0.012\pm 0.043$~\cite{BKPRD}).
This is because, although the latter has a smaller discretization error,
it used only on the three larger lattice spacings and so has a larger
truncation error.}

In summary, improving the statistical errors has brought to light
a difficulty in the continuum extrapolation that was previously
masked. Although we have not fully understood the $a^2$ dependence,
this has relatively little impact on our final result, because the
extrapolation is anchored by the result from the smallest lattice spacing,
a result which has not changed significantly.
Nevertheless, we intend to try more elaborate fits to improve our
understanding of the $a^2$ dependence.
We also need to revisit the sea-quark mass dependence, since the
lower value for $B_K$ on the F1 ensemble is now significantly
different from that on the F2 ensemble. This is in contrast to the very weak
sea-quark dependence on the coarse ensembles.
Our most important need, however, 
is to reduce the truncation error, which we aim to
do both using non-perturbative renormalization and two-loop matching.


\section{Acknowledgments}
C.~Jung is supported by the US DOE under contract DE-AC02-98CH10886.
The research of W.~Lee is supported by the Creative Research
Initiatives Program (3348-20090015) of the NRF grant funded by the
Korean government (MEST). 
W.~Lee would like to acknowledge the support from KISTI supercomputing
center through the strategic support program for the supercomputing
application research [No. KSC-2011-C3-03].
The work of S.~Sharpe is supported in part by the US DOE grant
no.~DE-FG02-96ER40956.
Computations were carried out in part on QCDOC computing facilities of
the USQCD Collaboration at Brookhaven National Lab, and on the DAVID
GPU clusters at Seoul National University. The USQCD Collaboration are
funded by the Office of Science of the U.S. Department of Energy.

\end{document}